\documentclass[graybox, secnum]{svmult}

\usepackage{mathptmx}       % selects Times Roman as basic font
\usepackage{helvet}         % selects Helvetica as sans-serif font
\usepackage{courier}        % selects Courier as typewriter font
\usepackage{type1cm}        % activate if the above 3 fonts are
\usepackage{makeidx}         % allows index generation
\usepackage{graphicx}        % standard LaTeX graphics tool
\usepackage{multicol}        % used for the two-column index
\usepackage[bottom]{footmisc}% places footnotes at page bottom
\usepackage{hyperref}        %for hyperlinks
\usepackage{soul}            % for high-lighting of text
\hypersetup{colorlinks=true,urlcolor=blue}
\usepackage[square,numbers]{natbib}
\makeindex             % used for the subject index
\graphicspath{{./}{figures/}}
\begin{document}
%\tableofcontents{}
\title*{Supernova remnants: Types and evolution}
\author{Aya Bamba\thanks{corresponding author} and Brian J. Williams}
\institute{Aya Bamba \at The University of Tokyo, Hongo 7-3-1, Bunkyo-ku, Tokyo, Japan 113-0033, \email{bamba@phys.s.u-tokyo.ac.jp}
\and Brian J. Williams \at NASA Goddard Space Flight Center, X-ray Astrophysics Laboratory, Greenbelt, MD 20771 \email{brian.j.williams@nasa.gov}}
\maketitle

%\tableofcontents{}

%
\abstract{
Although only a small fraction of stars end their lives as supernovae, all supernovae leave behind a supernova remnant (SNR), an expanding shock wave that interacts with the surrounding medium, heating the gas and seeding the cosmos with elements forged in the progenitor In this chapter, we introduce the basic properties of galactic and extragalactic SNRs (\S\ref{sec:SNR-intro}). We summarize how SNRs evolve throughout their life cycles over the course of $\sim 10^{6}$ yrs (\S\ref{sec:SNR-evolution}).
We discuss the various morphological types of SNRs and discuss the emission processes at various wavelengths.(\S\ref{sec:SNR-types}).
%Each chapter should be preceded by an abstract (about 250 words) that summarizes the content. The abstract will appear \textit{online} at \url{www.SpringerLink.com} and be available with unrestricted access. This allows unregistered users to read the abstract as a teaser for the complete chapter. Please do not include reference citations, cross-references or undefined abbreviations in the abstract.
}
\section{Keywords} 
Supernova remnants (SNRs) --- free expansion phase --- adiabatic expansion phase
--- shell-type SNR --- plerion-type SNR --- mixed morphology SNR
%Please provide keywords required to facilitate search of chapter on web; maximum 10 keywords.

\section{Introduction}
\label{sec:SNR-intro}

Supernova (SN) explosions are among the most energetic events in the universe since the Big Bang, releasing more energy ($\sim 10^{51} - 10^{53}$ ergs) than the Sun will release over its entire lifetime. They are the cataclysmic ends of certain types of stars, and are responsible for seeding the universe with the material necessary to form other stars, planets, and life itself. We owe our very existence to generations of stars that lived and died billions of years ago, before the formation of our Sun and solar system. The processes of stellar evolution continue to occur today, with typical galaxies like the Milky Way hosting a few supernovae per century, on average. The study of supernovae and the role they play in shaping the evolution of star systems and galaxies is truly an exploration of our own origins. Supernova remnants (SNRs), the expanding clouds of material that remain after the explosion, spread elements over volumes of thousands of cubic light-years, and heat the interstellar medium through fast shock waves generated by the ejecta from the star.

There are currently $\sim 300$ SNRs cataloged in our Galaxy \citep{green2019},
although radio surface brightness studies predict that
only half of SNRs have been identified \citep{case1998}.
Recent X-ray observations with good spatial resolution and large effective area allow us to detect many SNR samples 
in nearby galaxies as well, such as Magellanic Clouds\cite{maggi2016}, M31\cite{sasaki2012}, and M33\cite{garofali2017}.

%In this chapter, we will discuss the evolution of SNRs and their classifications.

\section{Evolution of supernova remnants}
\label{sec:SNR-evolution}

\begin{figure}[h]
\begin{center}
\includegraphics[width=0.8\textwidth]{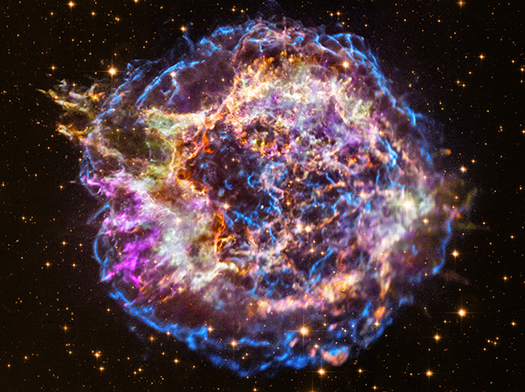}
\caption{Cassiopeia A, a young remnant in our Galaxy, as seen at X-ray energies with {\it Chandra}. Credit: NASA/CXC/SAO
\label{casa}}
\end{center}
\end{figure}

The general picture of an SNR is that of a cloud of expanding material ejected from the star, rich in heavy elements like O, Si, and Fe, that drive a shock wave into the interstellar medium (ISM). This shock waves heat the interstellar gas it encounters to millions of degrees, creating a highly-ionized plasma. Figure~\ref{casa} shows Cassiopeia A in X-rays, an example of a young ($\sim 350$ yr) SNR in our own galaxy. Although Cassiopeia A is known to have resulted from a core-collapse (CC) SN {\bf (see Chapter on SN)}, it is generally difficult to tell the type of SN only by looking at the remnant. SNRs remain visible for thousands, often tens or hundreds of thousands of years before dissipating their energy into the ISM. The life of a SNR can be thought of as consisting of four phases.

\subsection{Free expansion phase}

The ejected gas mainly determines the evolution of SNRs 
in the initial phase.
%{\bf The kinetic energy of the ejecta is roughly $10^{51}$~erg.}
%About 1\% of the explosion energy goes into the kinetic energy of the ejecta.
Since the pressure of surrounding matter is negligible in comparison, 
the exploded ejecta expands effectively without deceleration. As a result, this phase is called the ``free expansion'' phase.

{\bf The characteristic (or typical) speed of the ejecta expansion,}
%The speed of the ejecta expansion, 
$v_s$, is given by
\begin{equation}
    v_s = \sqrt{\frac{2E}{M_{ej}}}\ \ ,
\end{equation}
where $E$ and $M_{ej}$ are the kinetic energy of the explosion and the ejecta mass, respectively. With fiducial values of $E_{kin} = 1\times 10^{51}$~erg
and $M_{ej} = 1M_\odot$, a typical velocity of the ejecta is $\sim 10^4$~km~s$^{-1}$.
While this velocity for the ejecta is quite high,
 it is also a bulk velocity with very little randomization in the velocity of the various fluid elements
{\bf without deceleration (This is the difinition of ``free expansion")}. Thus, 
{\bf there is no heating and }the ejecta temperature is actually quite cold
{\bf in the beginning},
and the remnant is generally faint at all wavelengths. The exception is the radiation from nonthermal synchrotron emission
{\bf from high energy electrons accelerated in the shock 
(see Chapter ``acceleration").}

\begin{figure}[h]
\begin{center}
\includegraphics[width=0.4\textwidth]{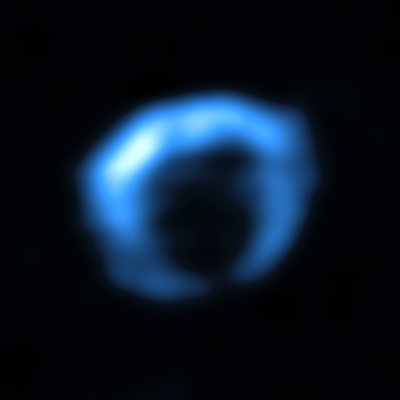}
\includegraphics[width=0.4\textwidth]{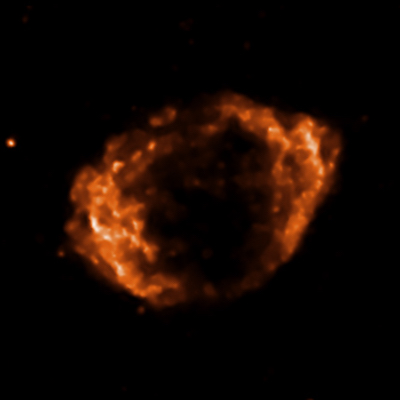}
\caption{{\it Left:} SNR G1.9+0.3, the youngest remnant in the Galaxy, seen in radio waves from the VLA in 1985. {\it Right:} The remnant, in 2007, seen in X-rays with Chandra. In the 22 years between these images, the remnant expanded substantially, allowing the time since the original supernova explosion (about 140 years) to be estimated. Credit: X-ray (NASA/CXC/NCSU/S.Reynolds et al.); Radio (NSF/NRAO/VLA/Cambridge/D.Green et al.).
\label{g1.9}}
\end{center}
\end{figure}

Virtually all SNRs emit in radio waves, where the emission is entirely nonthermal in origin, resulting from synchrotron radiation from relativistic electrons spiraling around magnetic fields {\bf in the shock}. Synchrotron emission is characterized by a featureless power-law spectrum, where the radio flux, $S_{\nu}$ is given by $S_{\nu} \propto \nu^{-\alpha}$, where $\nu$ is the frequency and $\alpha$ is the spectral index, which depends on the energy distribution of the electron population. 
{\bf For more detail, see Chapter ``acceleration".}
In Figure~\ref{g1.9}, we show an example of a remnant, G1.9$+$0.3, which is still in the free expansion phase \cite{reynolds08a}. The acceleration of electrons up to GeV energies is sufficient to produce radio synchrotron emission.

A small population of SNRs are apparently able to accelerate electrons to much higher energies ($> 10$ TeV), sufficient to produce synchrotron radiation at X-ray energies
\cite{koyama1995,aharonian2004}. G1.9+0.3 also falls into this category, and Figure~\ref{g1.9} also shows an image of the remnant that is essentially entirely from nonthermal synchrotron emission. For more details regarding the particle acceleration mechanisms, see the review by \cite{reynolds08b}.

%As the remnant evolves, accumulated interstellar medium is swept up by the forward shock. The front of discontinuity also makes a density shock,
%which runs toward the center of the SNR
%and is called as "reverse shock".
%The similarity solution is discussed by \citet{hamilton1984a,hamilton1984b}.

\subsection{Adiabatic expansion phase}

At the end of the free expansion phase, the expansion begins to decelerate due to the forward shock sweeping up a non-negligible amount of material in the ISM. An estimate for when this transition occurs can be obtained by calculating the radius, $R_{S}^{0}$, at which the swept-up mass equals the ejecta mass:

\begin{eqnarray}
    \frac{4}{3}\pi {R_s^0}^3 \rho_0 &=& M_{ej}\ \ , \\
    R_s^0 &=& 5.8 {\left(\frac{M_{ej}}{10M_\odot} \right)}^{1/3}
    {\left( \frac{n_0}{0.5\ {\rm cm}^{-3}} \ \right)}^{-1/3}\ ({\rm pc})\ \ ,
\end{eqnarray}

where $\rho_0$ and $n_0$ are mass and number densities
in the cgs unit.
Assuming typical values of supernova explosion and ISM,
it generally takes a few hundred years for the radius of the SNR to reach $R_s^0$ (though this can vary by orders of magnitude if the SN explodes into particularly low or high density regions of the ISM). 

This phase of deceleration of the forward shock wave also leads to the formation of a ``reverse shock,'' which propagates back into the ejecta towards the center of the SNR. Similarity solutions for the reverse shock are discussed in \citet{hamilton1984a,hamilton1984b}.

In this phase, the evolution of the SNR radius as a function of time can be written as
\begin{equation}
    R(t) \propto t^m\ \ ,
\end{equation}
where $m$ is the expansion parameter defined by 
\cite{hughes1999}.
At the end of the free expansion phase, $m \approx 1$, by definition, but during this second phase, $m$ gradually decreases to $m=0.4$ \cite{sedov1959}.

When the SNR shock starts decelerating, radiative cooling is still negligible
and the expansion is adiabatic. For the most simple case of a uniform ISM, the evolution can be written as the results of the similarity solution by \cite{sedov1959} (thus, this phase is often referred to as the ``Sedov phase").

Assuming that a large amount of energy is instantaneously released in a small volume,
expansion can be characterized {\bf by} only two parameters,
$\rho_0$ (the mass density of the ISM) and $E_{kin}$ (the kinetic energy of the explosion).
We can define a non-dimensional parameter $\zeta$ as
\begin{equation}
    \zeta = R_s{\left( \frac{Et^2}{\rho_0}\right)}^{-1/5}\ \ ,
\end{equation}
where $R_s$ and $t$ represent the radius and age of the SNR.
$R_s$ can be written as 
\begin{equation}
    R_s = \zeta_0{\left(\frac{Et^2}{\rho_0}\right)}^{1/5}\ \ ,
\end{equation}
where $\zeta_0$ is determined from the energy conservation equation. 
Using $\zeta_0 = 1.17$ for ideal gas ($\gamma = 5/3$)
\cite{landau1959},
$R_s$, the temperature $T$, and the shock velocity $v_s$ are written as
\begin{eqnarray}
R_s &=& 5.0{\left(\frac{E_{51}}{n_0} \right)}^{1/5}t_3{}^{2/5}\ \ {\rm (pc)}\ \ ,\\
T_s &=& 1.5\times 10^{10}\left(\frac{E_{51}}{n_0}\right)R_s^{-3}\ \ {\rm (K)}\\
&=& 1.2\times 10^8 \times {\left(\frac{E_{51}}{n_0}\right)}^{2/5}t_3{}^{-6/5}\ \ {\rm (K)}\ \ ,\\
v_s &=& \frac{dR_s}{dt} = 2.1\times 10^8 \times{\left(\frac{E_{51}}{n_0}\right)}^{1/5}t_3{}^{-3/5}\ \ {\rm (cm\ s^{-1})}\ \ ,
\end{eqnarray}
where $E_{51}$, $n_0$, and $t_3$ are the explosion energy in the unit of $10^{51}$~erg, the number density of the ambient gas in the unit of cm$^{-3}$, and the SNR age in the unit of $10^3$~years.
During the adiabatic expansion phase, $\sim 70\%$ of the initial explosion energy is converted into thermal energy of swept-up matter \cite{chevalier1974}.

\begin{figure}[h]
\begin{center}
\includegraphics[width=0.8\textwidth]{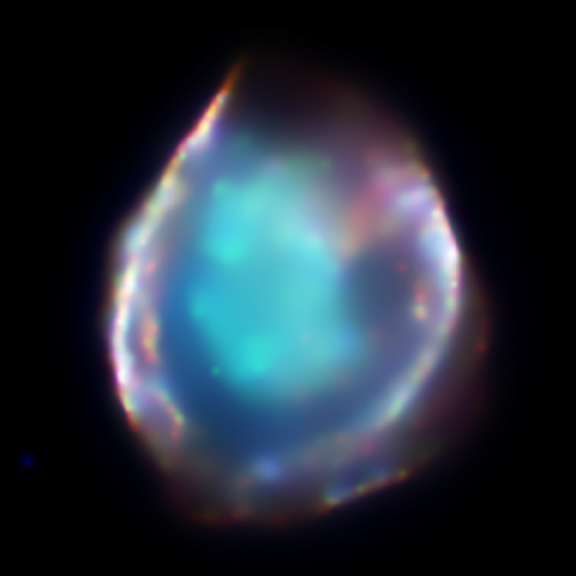}
\caption{SNR DEM L71, located in the LMC. This X-ray image shows a clear separation between the forward shocked material in the ring on the outside and the reverse-shocked material in blue on the inside. Credit: NASA/CXC/Rutgers/J.Hughes et al. 
\label{demL71}}
\end{center}
\end{figure}

Remnants in the adiabatic or Sedov expansion phase are typically bright at X-ray wavelengths, owing to both the hot ISM gas shocked by the forward shock and the hot ejecta shocked by the reverse shock. In Figure~\ref{demL71}, we show a canonical example of a remnant in this phase: DEM L71, a remnant in the LMC believed to be a few thousand years post-explosion \cite{hughes02,rakowski03}.

High-mass stars typically eject mass as stellar winds before their explosions, forming a circumstellar medium (CSM).
The winds make a density gradient around the progenitor star, making the deceleration of the shock slower. A modified similarity solution for this case is shown in \cite{truelove1999}.

These models introduced here ignored radiative cooling
under the assumption of an ideal gas. In the case that the system is not truly adiabatic, energy is robbed from the shock and $v_s$ becomes smaller.
Such a case can be caused by efficient acceleration of cosmic rays at the shock front \cite[for example]{berezhko2002}.

\subsection{Snowplow phase}

\begin{figure}[h]
\begin{center}
\includegraphics[width=0.8\textwidth]{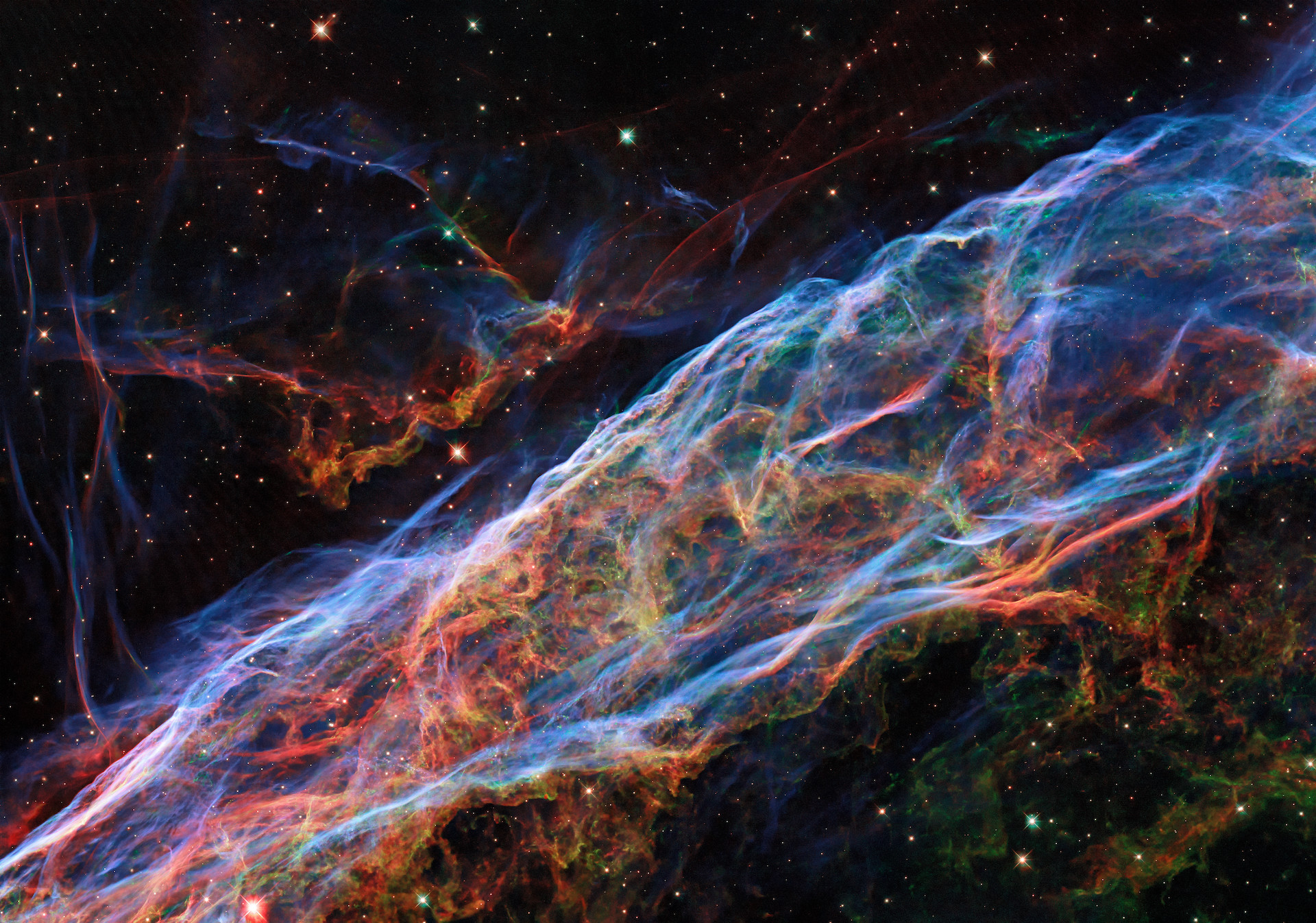}
\caption{A portion of the Cygnus Loop SNR, seen with {\it HST}. Narrowband images of [O III], [S II], and H$\alpha$, plus {\it V} and {\it I} continuum bands combine to make this image. Credit: NASA/ESA/Hubble Heritage Team. 
\label{cygnusloop}}
\end{center}
\end{figure}

When the shock velocity has decelerated to $\sim$200~km~s$^{-1}$, the temperature of the gas drops below $\sim 10^{6}$~K and radiative cooling begins to affect the dynamics of the shock evolution. Cooling of the gas is a runaway process, as the more it cools, the more the cooling rate increases. The density of the shock front becomes highest where the timescale of radiative cooling becomes shortest; as a result, a cool, dense shell is formed just behind the shock front,
whereas the inner gas still has high temperature and high pressure 
due to the low density. It is during this phase that the remnant becomes bright in optical emission from recombining atoms, most notably H$\alpha$, [S II], [N II], and [O III]. As an example of a remnant in this phase, we show an {\it HST} image of a portion of the Cygnus Loop in Figure~\ref{cygnusloop}. The Cygnus Loop is the remains of a SN believed to be 10,000-20,000 yr post explosion. 

The cool shell continues the expansion further collecting ISM gas in a manner reminiscent of a snowplow (hence the name "snowplow phase").
At first, the gas expands adiabatically, then $pV^{5/3}$ becomes a constant.
The shock expands with the time dependency of $R_s\propto t^{2/7}$.
As the temperature cools down further, the pressure can be ignored and the cool shell expands at a constant radial momentum with a time dependency of roughly $R_s \propto t^{1/4}$.

\subsection{Dissipation phase}

As the SNR gets even older, the expansion velocity becomes smaller.
When the shock slows down to a velocity comparable to {\bf the sound velocity},
%the proper motion of surrounding interstellar medium (10--20~km~s$^{-1}$),
the SNR blends into the interstellar medium.
This is called the ``Dissipation phase," and marks the end of the SNR phase.
The lifetime of typical SNRs is $\sim 10^6$~yrs, during which time they will expand to radii of dozens to hundreds of parsecs, filling a volume large enough to encompass hundreds to thousands of star systems.

\section{Types of supernova remnants}
\label{sec:SNR-types}

\subsection{Shell-type SNRs}

The majority of remnants are classified as ``shell-type" SNRs, where the forward shock is delineated by emission at wavelengths across the electro-magnetic spectrum. The remnants shown in Figures 1$-$4 are all shell-type SNRs. The physical processes responsible for emitting at various wavelengths differ. In radio regime, emission is dominated by synchrotron emission from energetic electrons (accelerated by the shock wave to GeV energies) spiraling around the turbulent magnetic field at the shock front. Historically, most SNRs have been discovered and categorized based on their radio emission \citep{green2019}.

At infrared wavelengths, young SNRs (those in the free expansion or adiabatic expansion phases) are generally dominated by thermal emission from warm dust grains; see \cite{dwek1992} for a review. Dust grains themselves do not generally feel the passage of the shock wave, but are suddenly immersed in the hot plasma behind the shock, where ion and electron temperatures can be tens of millions of degrees. Grains are collisionally heated by these particles, and reach temperatures of 50-200 K. Dust at these temperatures radiates in the mid-IR, with thermal spectra peaking in the 20-100 $\mu$m range. 

\begin{figure}[h]
\begin{center}
\includegraphics[width=0.3\textwidth]{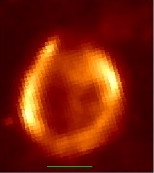}
\includegraphics[width=0.633\textwidth]{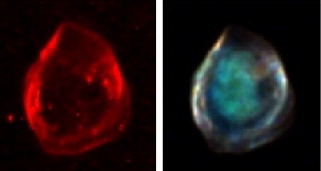}
\caption{{\it Left}: The {\it Spitzer} 24 $\mu$m emission from DEM L71. {\it Middle}: The remnant in H$\alpha$. {\it Right}: The remnant in X-rays. Figure modified from Figure 1 in \cite{borkowski2006}. 
\label{dust}}
\end{center}
\end{figure}

Generally speaking, IR emission from young shell-type SNRs only shows up in emission from the shell of the remnant. This emission results from warmed interstellar grains; searches for dust associated with the ejecta in SNRs has turned up very little \citep{borkowski2006,williams2006}. As an example, in Figure~\ref{dust}, we show the IR emission from DEM L71, the remnant whose X-ray emission is shown in Figure~\ref{demL71}.

At optical wavelengths, emission from the shell of a shell-type remnant can be seen in one of two ways. If there is at least partially neutral material present ahead of the shock (often taken as an indication that the object is the remnant of a Type Ia SN; see \cite{williams2011}), then charge exchange can take place in the post-shock environment giving rise to a strong H$\alpha$ component \citep{chevalier1980}, as shown in the middle panel of Figure~\ref{dust}. The temperatures in these remnants is too hot to allow recombination lines in other elements. In older remnants where the temperatures have cooled significantly, recombination lines from various elements begin to dominate (see Section~\ref{sec:SNR-evolution}).

At X-ray wavelengths, shell-type SNRs may show emission from either or both of the forward-shocked ISM/CSM and the reverse-shocked ejecta, depending on the evolutionary state of the remnant. The physical processes causing the X-ray emission can be either thermal, non-thermal, or both. The thermal emission is generally a combination of both Bremmsstrahlung continuum produced when hot ions and electrons interact, as well as line emission from highly-ionized atoms of various metals, such as O, Fe, Ne, Si, and S. Nonthermal emission in SNRs arises from synchrotron emission identical to that seen in radio waves, but from much more energetic ($\sim 10-100$ TeV) electrons. Because electrons can only be accelerated to energies this high very close to the blast wave, synchrotron emission is most often seen from the outer edges of the shell in shell-type SNRs \cite{bamba2003,bamba2005}. 

Shell-type remnants with sufficient spatial extent can even be identified in gamma-rays. For example, \cite{katagiri2011} used {\it Fermi} observations to show that the Cygnus Loop is spatially-resolved into a shell in the 0.5-10 GeV energy range. The various physical processes involved in gamma-ray emission from SNRs is beyond the scope of this chapter; for a thorough review, see \cite{reynolds2008}.

\begin{figure}[h]
\begin{center}
    \includegraphics[width=0.8\textwidth]{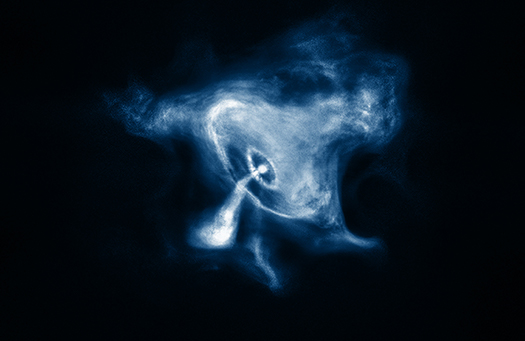}
    \caption{X-ray image of the Crab nebula taken with {\it Chandra}. (credit: NASA/CXC/SAO)}
    \label{fig:crab}
\end{center}
\end{figure}

\subsection{Plerion-type SNRs}

Young neutron stars born {\bf in} core-collapse SNe often form pulsar wind nebulae.
Pulsar wind nebulae emit bright nonthermal emission from radio to very high energy gamma-rays,
via synchrotron or inverse Compton processes.
These are referred to as ``plerion-type,'' or ``plerionic" SNRs \cite{weiler1978}.

The typical plerion-type is the Crab nebula.
It is the remnant of SN~1054, and is now one of the brightest celestial sources at most wavelengths.
Figure~\ref{fig:crab} shows the X-ray image of the Crab nebula taken by Chandra.
Jets extend out to northwest and southeast from the central source (the pulsar),
while two torii are seen emanating outward from the center.
The inner radius of the torus is 0.01--0.1~pc,
which corresponds the shock of the relativistic winds from the pulsar terminated by ambient material
\cite{rees1974,kennel1984,bamba2010}.
For more details, please see the chapter of this work concerning PWNe.

\begin{figure}[h]
\begin{center}
    \includegraphics[width=0.6\textwidth]{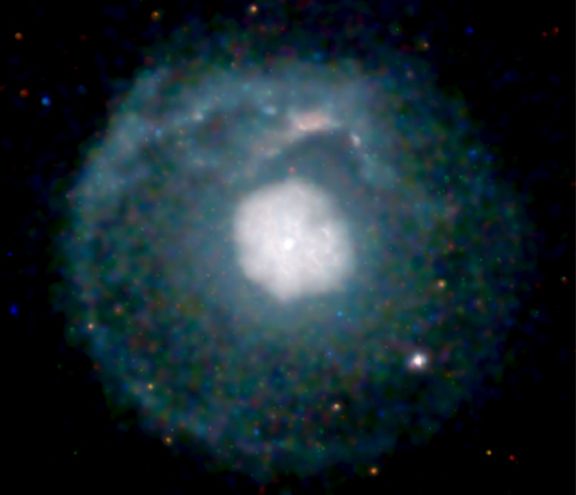}
    \caption{X-ray image of the plerion-type SNR, G21.5$-$0.9 taken with Chandra. (credit: Credit: Heather Matheson \ Samar Safi-Harb (Univ. Manitoba), CXC, NASA)}
    \label{fig:g21.5}
\end{center}
\end{figure}

Some plerion-type SNRs are surrounded by shells of emission resulting from the forward shock interacting with the surrounding medium. Such remnants are often called ``composite type." One of the typical composite type SNRs is 
G21.5$-$0.9 \cite{slane2000}.
Figure~\ref{fig:g21.5} is the X-ray image of G21.5$-$0.9. One can see both the central PWN and the outer shell.

The Crab nebula is a case of a remnant without its shell; it is still an open issue why there is no detected shell surrounding the Crab nebula.
Recent observations by the {\it Hitomi} satellite showed
that the thermal X-ray emission around the Crab nebula is significantly fainter than other young SNRs
\cite{hitomi2018}.
The total ejecta mass is estimated to be less than $\sim 1M_\odot$,
implying that the progenitor explosion was relatively low energy.
One possibility is that
the Crab nebula is a remnant of an electron capture supernova \cite{miyaji1980}.
More detailed observations for the Crab and more samples are needed
to understand the origin of plerion-type SNRs.

\subsection{Mixed-Morphology SNRs}

Some relatively evolved SNRs show different morphology 
in radio and X-ray bands;
shell-like structure in radio,
whereas the centrally-filed in X-rays.
They are called ``Mixed-Morphology SNRs" or MM SNRs
\citep{rho1998}.
The X-rays are thermal, thus it is not due to its central pulsar and/or pulsar wine nebula
like plerion cases.

The mechanism by which mixed-morphology remnants are formed is still unclear.
Several hypotheses have been discussed to explain how
a remnant with a radio shell can appear centrally
filled in X-rays.
The first possibility is ``fossil emission" of the remnant \cite{seward1985};
the outer shell cools down due to the expansion and radiative cooling,
and as a result, the shock temperature becomes too low to emit thermal X-rays.
Synchrotron emission from the accelerated electrons in the shock disappear from the high energy side due to the synchrotron cooling.
Thus the shells emitting synchrotron disappear within a few thousand years in the X-ray band, whereas they remain for more than ten thousand years in the radio band.

Another possibility is due to ``evaporated cloud";
the enhanced interior X-ray emission could arise from gas evaporated from clouds \cite{white1991}
in the cases that shocks of SNRs propagates through a cloudy interstellar medium.
Dense and small size clouds can pass the shock
and later evaporate and are heated,
and as a result makes thermal X-ray emission
inside the shells.

\section{Conclusions}

Supernova remnants are an incredibly diverse class of objects. From simply a purely morphological standpoint, their are influenced by a number of factors, including mass loss from the progenitor system, the energy and isotropy of the explosion itself, and the structure of the surrounding ISM/CSM. The radiation they emit also varies from source to source: some remnants are bright in radio waves, some in optical, some in X-rays. In general, the remnant's state of dynamical evolution (which can be roughly thought of as the product of age and the density being encountered by the forward shock) is the most important factor for determining the spectral energy distribution for a particular SNR. Other chapters within this work will go into more detail on many aspects of SNR study.


\begin{thebibliography}{99.}

\bibitem[Green(2019)]{green2019}
Green, D.~A.\ 2019, Journal of Astrophysics and Astronomy, 40, 36. doi:10.1007/s12036-019-9601-6

\bibitem[Case \& Bhattacharya(1998)]{case1998}
Case, G.~L. \& Bhattacharya, D.\ 1998, THe Astrophysical J., 504, 761. doi:10.1086/306089

\bibitem[Maggi et al.(2016)]{maggi2016}
Maggi, P., Haberl, F., Kavanagh, P.~J., et al.\ 2016, A\&A, 585, A162. doi:10.1051/0004-6361/201526932


\bibitem[Sasaki et al.(2012)]{sasaki2012}
Sasaki, M., Pietsch, W., Haberl, F., et al.\ 2012, A\&A, 544, A144. doi:10.1051/0004-6361/201219025

\bibitem[Garofali et al.(2017)]{garofali2017}
Garofali, K., Williams, B.~F., Plucinsky, P.~P., et al.\ 2017, MNRAS, 472, 308. doi:10.1093/mnras/stx1905

\bibitem[Reynolds et al.(2008a)]{reynolds08a} Reynolds, S.~P., et al. 2008, ApJ, 680, 41

\bibitem[Koyama et al.(1995)]{koyama1995}
Koyama, K., Petre, R., Gotthelf, E.~V., et al.\ 1995, Nature, 378, 255

\bibitem[Aharonian et al.(2004)]{aharonian2004}
Aharonian, F.~A., Akhperjanian, A.~G., Aye, K.-M., et al.\ 2004, Nature, 432, 75

\bibitem[Reynolds et al.(2008b)]{reynolds08b} Reynolds, S.~P. 2008, ARAA, 46, 89

\bibitem[Hamilton \& Sarazin(1984a)]{hamilton1984a}
Hamilton, A.~J.~S. \& Sarazin, C.~L.\ 1984, The Astrophyical J., 287, 282. doi:10.1086/162687

\bibitem[Hamilton \& Sarazin(1984b)]{hamilton1984b}
Hamilton, A.~J.~S. \& Sarazin, C.~L.\ 1984, The Astrophysical J., 281, 682. doi:10.1086/162145

\bibitem[Hughes(1999)]{hughes1999}
Hughes, J.~P.\ 1999, The Astrophysical J., 527, 298. doi:10.1086/308082

\bibitem[Sedov(1959)]{sedov1959}
Sedov, L.~I.\ 1959, Similarity and Dimensional Methods in Mechanics, New York: Academic Press, 1959

\bibitem[Landau \& Lifshitz(1959)]{landau1959}
Landau, L.~D. \& Lifshitz, E.~M.\ 1959, Course of theoretical physics, Oxford: Pergamon Press, 1959

\bibitem[Chevalier(1974)]{chevalier1974}
Chevalier, R.~A.\ 1974, The Astrophysical J., 188, 501. doi:10.1086/152740

\bibitem[Hughes et al.(2002)]{hughes02} Hughes, J.~P., Ghavamian, P., Rakowski, C.~E., \& Slane, P.~O.\ 2003, ApJ, 582, 95

\bibitem[Rakowski et al.(2003)]{rakowski03} Rakowski, C.~E., Ghavamian, P., \& Hughes, J.~P.\ 2003, ApJ, 590, 846

\bibitem[Truelove \& McKee(1999)]{truelove1999} Truelove, J.~K. \& McKee, C.~F.\ 1999, The Astrophysical J. Suppliment, 120, 299. doi:10.1086/313176

\bibitem[Berezhko et al.(2002)]{berezhko2002} Berezhko, E.~G., Petukhov, S.~I., \& Taneev, S.~N.\ 2002, Astronomy Letters, 28, 632. doi:10.1134/1.1505508

\bibitem[Rees \& Gunn(1974)]{rees1974}
Rees, M.~J. \& Gunn, J.~E.\ 1974, MNRAS, 167, 1

\bibitem[Kennel \& Coroniti(1984)]{kennel1984}
Kennel, C.~F. \& Coroniti, F.~V.\ 1984, ApJ, 283, 694

\bibitem[Bamba et al.(2010)]{bamba2010}
Bamba, A., Mori, K., \& Shibata, S.\ 2010, ApJ, 709, 507

\bibitem[Slane et al.(2000)]{slane2000}
Slane, P., Chen, Y., Schulz, N.~S., et al.\ 2000, ApJL, 533, L29

\bibitem[Hitomi Collaboration(2018)]{hitomi2018}
Hitomi Collaboration, Aharonian, F., Akamatsu, H., et al.\ 2018, PASJ, 70

\bibitem[Rho \& Petre(1998)]{rho1998}
Rho, J. \& Petre, R.\ 1998, ApJL, 503, L167. doi:10.1086/311538

\bibitem[Dwek \& Arendt(1992)]{dwek1992}
Dwek, E., \& Arendt, R.\ 1992, ARA\&A, 30, 11

\bibitem[Borkowski et al.(2006)]{borkowski2006}
Borkowski, K., et al. 2006, ApJ, 642, 141

\bibitem[Williams et al.(2006)]{williams2006}
Williams, B.J., et al. 2006, ApJ, 652, 33

\bibitem[Williams et al.(2011)]{williams2011}
Williams, B.J., et al. 2011, ApJ, 741, 96

\bibitem[Chevalier et al.(1980)]{chevalier1980}
Chevalier, R.A., Kirshner, R.P., \& Raymond, J.C. 1980, ApJ, 235, 186

\bibitem[Bamba et al.(2003)]{bamba2003}
Bamba, A., Yamazaki, R., Ueno, M., et al.\ 2003, ApJ, 589, 827

\bibitem[Bamba et al.(2005)]{bamba2005}
Bamba, A., Yamazaki, R., Yoshida, T., et al.\ 2005, ApJ, 621, 793



\bibitem[Katagiri et al.(2011)]{katagiri2011}
Katagiri, H., et al. 2011, ApJ, 741, 44

\bibitem[Reynolds(2008)]{reynolds2008}
Reynolds, S.P. 2008, ARA\&A, 46, 89

\bibitem[Weiler \& Panagia(1978)]{weiler1978}
Weiler, K.W. \& Panagia, N. 1978, A\&A, 70, 419

\bibitem[Miyaji et al.(1980)]{miyaji1980}
Miyaji, S., Nomoto, K., Yokoi, K., \& Sugimoto, D. 1980, PASJ, 32, 303

\bibitem[Seward(1985)]{seward1985}
Seward, F.~D.\ 1985, Comments on Astrophysics, 11, 15

\bibitem[White \& Long(1991)]{white1991}
White, R.~L. \& Long, K.~S.\ 1991, ApJ, 373, 543

% Book Chapter
%\bibitem{basic-contrib} Brown B, Aaron M (2001) The politics of nature. In: Smith J (ed) The rise of modern genomics, 3rd edn. Wiley, New York, p 234--295  
%
% Online Document
%\bibitem{basic-online} Dod J (1999) Effective Substances. In: The dictionary of substances and their effects. Royal Society of Chemistry. Available via DIALOG. \\
%\url{http://www.rsc.org/dose/title of subordinate document. Cited 15 Jan 1999}
%
% Journal article by DOI
%\bibitem{basic-DOI} Slifka MK, Whitton JL (2000) Clinical implications of dysregulated cytokine production. J Mol Med, doi: 10.1007/s001090000086
%
% Journal article
%\bibitem{basic-journal} Smith J, Jones M Jr, Houghton L et al (1999) Future of health insurance. N Engl J Med 965:325--329
%
% Monograph
%\bibitem{basic-mono} South J, Blass B (2001) The future of modern genomics. Blackwell, London 
%
\end{thebibliography}
\end{document}